\pdfoutput=1
\documentclass[conference]{IEEEtran}
\IEEEoverridecommandlockouts
\usepackage{amsmath,amssymb,amsbsy,amsfonts,balance}
\usepackage{graphicx}
\usepackage{epstopdf}
\usepackage{braket}
\usepackage{bm}
\usepackage{numprint}
\usepackage{float}
\usepackage[utf8]{inputenc}
\usepackage{xr}
\usepackage{amsthm}
\theoremstyle{definitions}

\usepackage{dsfont}
\usepackage{calc}
\usepackage{seqsplit}
\usepackage{stfloats}
\usepackage{hyperref}
\allowdisplaybreaks
\begin{document}

\title{On A Moving Average With Internal Degrees of Freedom}

\author{
\IEEEauthorblockN{Linda {Boudjemila}}
\IEEEauthorblockA{
\textit{\it University of Science} \\
\textit{\it and Technology Houari}\\
\textit{\it Boumedien Bab Ezzouar}\\
Algiers, Algeria, 16111 \\
lariessai21@gmail.com
}
\and
\IEEEauthorblockN{Alexander {Bobyl}}
\IEEEauthorblockA{
\textit{Ioffe Institute} \\
Polytechnicheskaya 26 \\
St. Petersburg, \\
Russia, 194021 \\
bobyl@theory.ioffe.ru
}
\and
\IEEEauthorblockN{Vadim {Davydov}}
\IEEEauthorblockA{
\textit{Peter the Great St.Petersburg} \\
\textit{Polytechnic University}\\
Polytechnicheskaya 29 \\
St. Petersburg  \\
Russia, 195251 \\
davydov\_vadim66@mail.ru
}
\and
\IEEEauthorblockN{Vladislav {Malyshkin}}
\IEEEauthorblockA{
\textit{Ioffe Institute} \\
Polytechnicheskaya 26 \\
St. Petersburg, \\
Russia, 194021 \\
{malyshki@ton.ioffe.ru}
}
}
\maketitle

\begin{abstract}
A new type of moving average is developed.
Whereas a regular moving average (e.g. of price) has
a built-in internal time scale (time--window, exponential weight, etc.),
the moving average developed in this paper has the
weight as the product of a polynomial by window factor.
The polynomial is the square of a wavefunction
obtained from an eigenproblem corresponding to
other observable (e.g. execution flow $I=dV/dt$, the number of shares
traded per unit time).
This allows to obtain an
immediate ``switch'' without lagging typical for regular moving average.
\end{abstract}

\begin{IEEEkeywords}
Radon Nikodym, fat tails, non--Gaussian noise
\end{IEEEkeywords}



\section{\label{intro}Introduction}
In the modern world, when conducting experiments and measurements,
data processing is of critical importance
\cite{gelgor2020flexible,petrov2020elimination,nguyen2019optimization,Semenov1991}.
To solve various problems, the data obtained as a result of measurements or calculations are grouped and further processed
\cite{ponomareva2021spectroscopic,davydov2018computer,davydov2019parameter,boudjemila2020electrical}.
One of these options is timeseries.
Timeserie data is widely available and used.
A typical problem in data analysis to study
timeserie
properties
and, possibly, trying to build a predictive model.
Built model are typically of autoregression,
Kalman filter\cite{kalman1960new,shrivastava2019overview,dhaya2021analysis,farahi2020probabilistic,ullah2019localization},
hidden Markov model\cite{rabiner1989tutorial,tran2021exponential,cheng2021asynchronous,kuppusamy2021asynchronous,wang2021asynchronous,dayanik2022detection,khalifa2021review},
and others\cite{einicke2012smoothing}.
In this work we consider timeserie data
of $M$ past observations, only past observations $l=1\dots M$
are available. The value of $M$ increases with every new observations coming.
A typical data of this type is timeserie data of financial transactions.
The most commonly available data is
(Time, Execution Price, Shares Traded) sequence of transactions
that we will be using for demonstration.
The problem of moving average\cite{gu2010detrending,van2019moving,singh2020prediction,mabude2021generally,yang2021least,lai2020use,kim2021dc,knoth2021impracticality} is considered.
Whereas typically considered moving average has
a built-in internal time scale (time--window, exponential weight, etc.),
the moving average developed in this paper has the
weight $d\mu=\psi^2(x(t))\omega(t)dt$ with
$\psi(x)$ being a polynomial obtained from eigenvalue problem
corresponding to a spike in other observable, such as execution flow $I=dV/dt$.
This allows to obtain an
immediate ``switch'' without lagging.
In previous work \cite{MalMuseScalp,malyshkin2022market} the ``switch'' was obtained
by applying an advanced technique of secondary sampling,
where a \textsl{calculated} value was used as it were
a new observable for sampling.
In this work we have implemented ``switching''
regime \textsl{without} using secondary sampling and generalized previous results
to be applied to a regular moving average. The main new result is
the developments 
of a new concept ---
a moving average with internal degrees of freedom.

\section{\label{Basis}Basis Selection and Integration Measure}
Introduce
the moments of basis polynomials $Q_m(x)$ (a polynomial of order $m$)
and integration measure $dt\,\omega(t)$ having
the support $t\in [-\infty\dots t_{now}]$:
\begin{align}
  \Braket{Q_m f}&=\int\limits_{-\infty}^{t_{now}} dt\,\omega(t)Q_m(x(t)) f(t)
  \label{momsDefinition}
\end{align}
The
$\omega(t)$ is decaying exponent
and
$x(t)$ is either linear or exponential function:
\begin{align}
  \omega(t)&=\exp\left(-(t_{now}-t)/\tau\right)
  \label{Wbasis} \\
  x(t)&=
  \begin{cases}
    (t-t_{now})/\tau & \text{Laguerre basis} \\
    \exp\left(-(t_{now}-t)/\tau\right) & \text{shifted Legendre basis}
  \end{cases}
  \label{Xbasis}
\end{align}
The moments of  $f$ are usually obtained
from direct sampling of all available observations $l=1\dots M$
in a timeserie:
\begin{align}
  \Braket{Q_m f}&=\sum_{l=1}^{M} f(t_l) Q_m(x(t_l)) \omega(t_l)
  \left[t_l-t_{l-1}\right]
\label{FSampleAll}
\end{align}
the moments of a derivative  $df/dt$
can also be obtained from direct sampling:
\begin{align}
  \Braket{Q_m \frac{df}{dt}}&=\sum_{l=1}^{M}
   Q_m(x(t_l)) \omega(t_l) \left[f(t_l)-f(t_{l-1})\right]
\label{dFSampleAll}
\end{align}
Given a good choice of basis polynomials:
\begin{align}
  Q_m(x)&=
  \begin{cases}
    L_m(-x) & \text{Laguerre basis} \\
    P_m(2x-1) & \text{shifted Legendre basis}
  \end{cases}
  \label{basesPol}
\end{align}
one can calculate (with double precision arithmetic) the moments to a very high order $m\lesssim 50$ (limited by the divergence of $c_m^{jk}$ multiplication coefficinets
(\ref{multiplicationOperator}))
in Laguerre basis and $m\lesssim 150$ (limited by poorly conditioned matrices)
in shifted Legendre basis;
Chebyshev polynomials $T_m(2x-1)$ also provide very stable calculations in shifted Legendre basis
(Chebyshev polynomials have perfectly stable multiplication:
all $c_m^{jk}=0$ except $c_{j-k}^{jk}=c_{j+k}^{jk}=0.5$, $j\ge k$).
All obtained results are invariant with respect
to basis choice,  $Q_m(x)=x^m$ and the ones from (\ref{basesPol})
give \textsl{identical} results, but numerical stability
can be drastically different\cite{beckermann1996numerical,2015arXiv151005510G}.

\section{\label{AverAndPsi}On Averaging Of Past Observations}

\begin{figure}[t]
  \includegraphics[width=8cm]{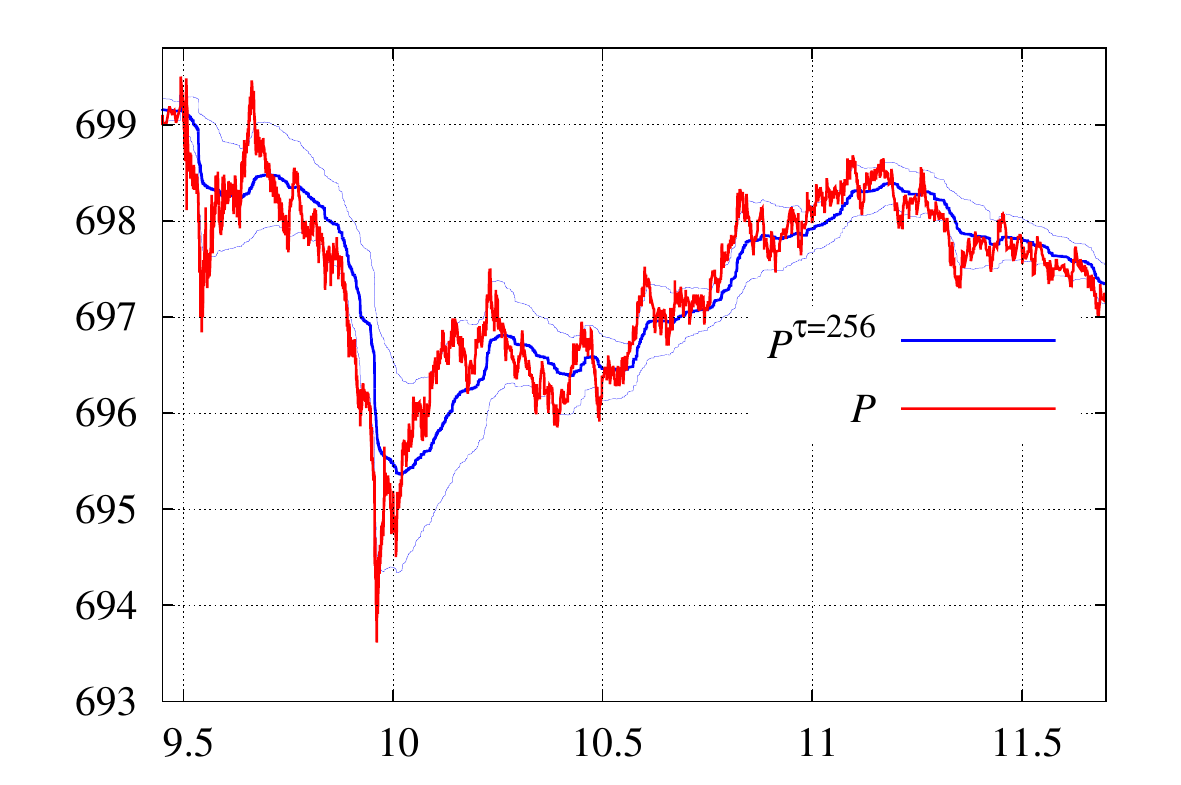}
  \caption{\label{MovingAveragePlot}
    An example of regular exponential moving average
    with $\tau=256$s.
    Standard deviation is also calculated with the same $\tau$ and
    moving average $\pm$ standard deviation is plotted as a thin line in the same color. As $\tau$ increases -- the moving average ``shifts to the right''
    ($\tau$-proportional time delay, lagging indicator).
 The data is for the AAPL stock on September, 20, 2012.
  }
\end{figure}

Consider familiar demonstration with
a moving average. Let $P^{\tau}$ be a regular exponential moving average.
The average $\Braket{\cdot}$
is calculated with the weight (\ref{Wbasis}):
\begin{align}
  P^{\tau}(t_{now})&=\frac{\Braket{pI}}{\Braket{I}}
  =\frac{\Braket{Q_0pI}}{\Braket{Q_0I}}
  =\frac{\int_{-\infty}^{t_{now}} dV\,\omega(t)p(t)}{\int_{-\infty}^{t_{now}} dV\,\omega(t)}
  \label{pmovingaver}
\end{align}
The averaging $d\mu=\omega(t)dt$ takes place between the past and $t_{now}$
using exponentially decaying weight $\omega(t)=\exp\left(-(t_{now}-t)/\tau\right)$.
With $\tau$ increase, the contributing to integral interval becomes larger
and moving average ``shifts to the right'' ($\tau$-proportional time delay, lagging indicator).
The (\ref{pmovingaver}) has no single parameter that can ``adjust''
the time scale,  see Fig. \ref{MovingAveragePlot}
for a demonstration.
From this demonstration it is clear that all moving average dependencies
are smooth, with the time scale of $\tau$.

Consider a different approach.
Introduce a wavefunction $\psi(x)$ as a linear combination
of basis function $Q_k(x)$:
\begin{eqnarray}
  \psi(x)&=&\sum\limits_{k=0}^{n-1} \alpha_k Q_k(x) \label{psiintr}
\end{eqnarray}
Then an observable market--related value
$f$, corresponding to the probability density  $\psi^2(x)$,
is calculated by averaging timeserie sample
with the weight $d\mu=\psi^2(x(t))\omega(t)dt$;
the expression corresponds to an estimation of Radon--Nikodym
derivative\cite{malyshkin2019radonnikodym}.
\begin{eqnarray}
  f_{\psi}&=&\frac{\Braket{\psi|f|\psi}}{\Braket{\psi|\psi}}
  \label{faver} \\
  f_{\psi}&=&\frac{\sum\limits_{j,k=0}^{n-1}\alpha_j\Braket{Q_j|f|Q_k}\alpha_k}{\sum\limits_{j,k=0}^{n-1}\alpha_j\Braket{Q_j|Q_k}\alpha_k}
  \label{faverexpand}
\end{eqnarray}
For averages we use
\href{https://en.wikipedia.org/wiki/Bra%E2%80%93ket_notation}{bra--ket notation}
  by
  \href{https://en.wikipedia.org/wiki/Paul_Dirac}{Paul Dirac}:
  $\Bra{\psi}$ and $\Ket{\psi}$.
The (\ref{faver}) is plain ratio of two moving averages,
but the weight is not regular decaying exponent $\omega(t)$
from (\ref{Wbasis}),
but exponent multiplied by wavefunction squared as $d\mu=\psi^2(x(t))\omega(t)dt$,
the $\psi^2(x)$ defines how to average a timeserie sample.
Any $\psi(x)$ function is defined by $n$ coefficients $\alpha_k$,
the value of an observable variable $f$ in $\psi(x)$ state
is a ratio of
two quadratic forms on $\alpha_k$ (\ref{faverexpand}).
Regular moving average (\ref{pmovingaver}) corresponds to $\psi(x)=const$.
Whereas typical approaches (Fourier, least squares, linear regression,
wavelets, etc.) deals with vector of moments $\Braket{Q_m f}$, $m=0\dots n-1$,
the quadratic forms ratio (\ref{faverexpand})  operates with matrices
$\Braket{Q_j| f|Q_k}$ and $\Braket{Q_j|Q_k}$.
The matrices can be obtained from $\Braket{Q_m f}$, $m=0\dots 2n-2$,
moments using
multiplication operator:
\begin{align}
  Q_j Q_k&=\sum_{m=0}^{j+k}c_m^{jk}Q_m
  \label{multiplicationOperator}
\end{align}
The main idea of \cite{2015arXiv151005510G}
is to consider a wavefunction (\ref{psiintr})
then to construct (\ref{faverexpand}) quadratic forms ratio.
A
\href{https://en.wikipedia.org/wiki/Eigendecomposition_of_a_matrix\#Generalized_eigenvalue_problem}{generalized eigenvalue problem}
is then considered with the two matrices from (\ref{faverexpand}).

We established,
that execution flow $I=dV/dt$ (the number of shares traded per unit time),
not trading volume $V$ (the number of shares traded),
is the driving force of the market: asset price is
much more sensitive to execution flow $I$ (dynamic impact),
rather than to traded volume $V$ (regular impact).
This corresponds to the matrices
$\Braket{Q_j|I|Q_k}$ and $\Braket{Q_j|Q_k}$.
These two matrices are volume- and time- averaged products of two basis functions.
Generalized eigenvalue problem for operator $I=dV/dt$
is then:
\begin{align}
&\Ket{I\middle|\psi^{[i]}}=\lambda^{[i]}\Ket{\psi^{[i]}} \label{GEVdef} \\
&\sum\limits_{k=0}^{n-1} \Braket{Q_j|I|Q_k} \alpha^{[i]}_k =
  \lambda^{[i]} \sum\limits_{k=0}^{n-1} \Braket{ Q_j|Q_k} \alpha^{[i]}_k
  \label{GEV} \\
 &\psi^{[i]}(x)=\sum\limits_{k=0}^{n-1} \alpha^{[i]}_k Q_k(x)
 \label{psiC}
\end{align}
Our analysis
shows that among the states $\Ket{\psi^{[i]}}$ of
the problem (\ref{GEVdef})
the state corresponding
to the maximal eigenvalue among all $\lambda^{[i]}$, $i=0\dots n-1$,
is the most important for market dynamics.
Consider various observable characteristics
in this state $\Ket{\psi^{[IH]}}$.

\begin{figure}[t]
  \includegraphics[width=8cm]{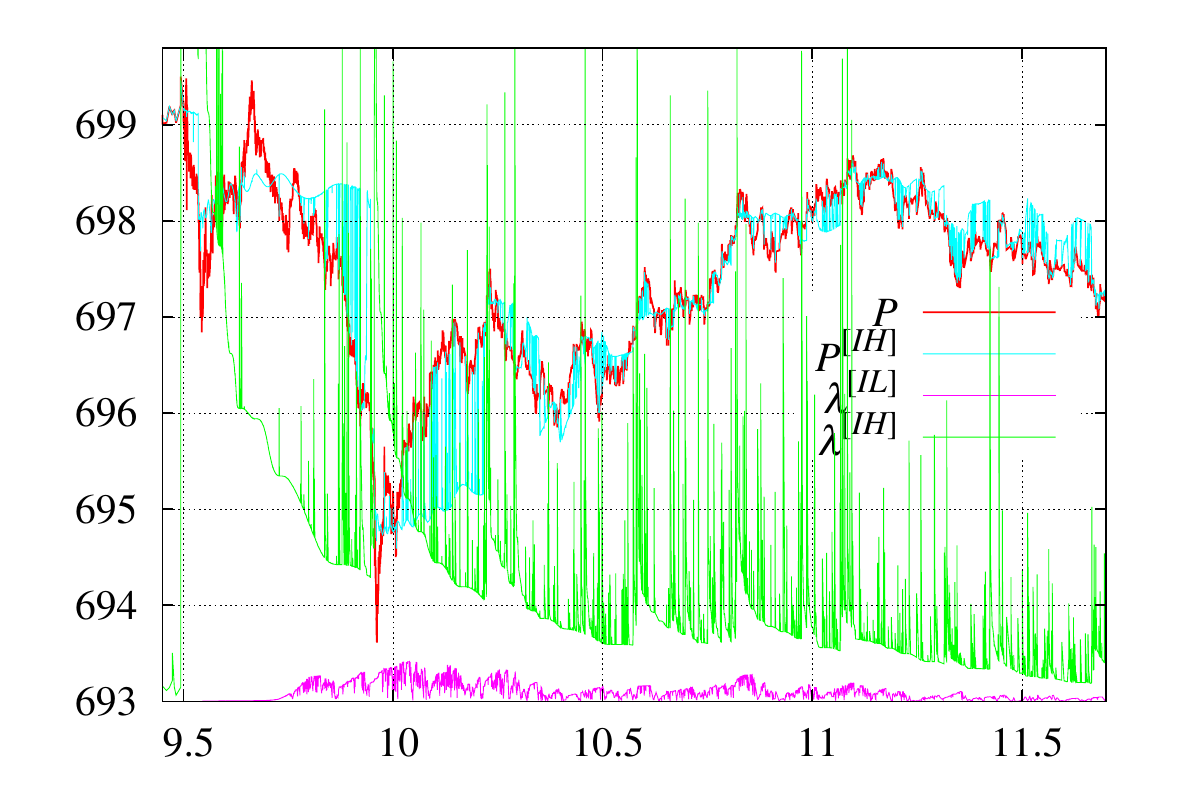}
  \caption{\label{ExampleIPsiH}
    Price $P$, price $P^{[IH]}$ (\ref{PIHGEV}),
    and maximal/minimal eigenvalues of (\ref{GEVdef})
    for
 the AAPL stock on September, 20, 2012.
 The calculations in shifted Legendre basis with $n=12$ and $\tau$=256sec.
  The execution flow  eigenvalues are scaled and shifted to $693$ to fit the chart.
  }
\end{figure}

In Fig. \ref{ExampleIPsiH} a demonstration of
several observables is presented:
the price in $\Ket{\psi^{[IH]}}$ state (\ref{PIHGEV}),
maximal eigenvalue $\lambda^{[IH]}$ of (\ref{GEVdef})   problem,
and minimal eigenvalue $\lambda^{[IL]}$ (for completeness).
\begin{align}
  P^{[IH]}&=\frac{\Braket{\psi^{[IH]}|pI|\psi^{[IH]}}}{\Braket{\psi^{[IH]}|I|\psi^{[IH]}}} \label{PIHGEV}
\end{align}
From these observable one can clearly see
that singularities in $I$ cause singularities in price,
and that a change in $\Ket{\psi^{[IH]}}$ localization causes
an immediate ``switch'' in an observable.
This switch is caused by the presence
of $n-1$ internal degrees of freedom
$\alpha_k$ ($n$ coefficients (\ref{psiC}), one less due to
normalizing $1=\Braket{\psi|\psi}$).
Such a ``switch'' is not possible in regular moving average
(\ref{pmovingaver})
since it has no any internal degree of freedom,
hence, all regular moving average dependencies are smooth.

\section{\label{LocalizationChanges}On Localization Changes in $\Ket{\psi^{[IH]}}$}
Considered above
the state  $\Ket{\psi^{[IH]}}$
maximizes the number of shares traded per unit
time on past observations sample; it
determines the time scale.
Let us consider in this state not the price and execution flow
as we studied before, but simply time distance to ``now'' $T^{[IH]}$:
\begin{align}
  T^{[IH]}&=\frac{\Braket{\psi^{[IH]}|(t_{now}-t)I|\psi^{[IH]}}}{\Braket{\psi^{[IH]}|I|\psi^{[IH]}}} \label{TIHGEV} \\
  T^{\tau}&=\frac{\Braket{(t_{now}-t)I}}{\Braket{I}}
  \label{TregularMovingAverage}
\end{align}
to compare it with regular moving average $T^{\tau}$ (were it
an integral over $dt$ the $T^{\tau}$ would be a constant;
the invergation with $dV$ make it tracking the spikes in $I$).
As all the values of time (future and past)
are known, the (\ref{TIHGEV}) carry information
about $\Ket{\psi^{[IH]}}$ localization.
When the value is small -- a large $dV/dt$ spike event happened
very recently. When it is large -- a large spike happened a substantial time ago,
the value is an information when a large spike in $dV/dt$ took place.
\begin{figure}[t]


  \includegraphics[width=8cm]{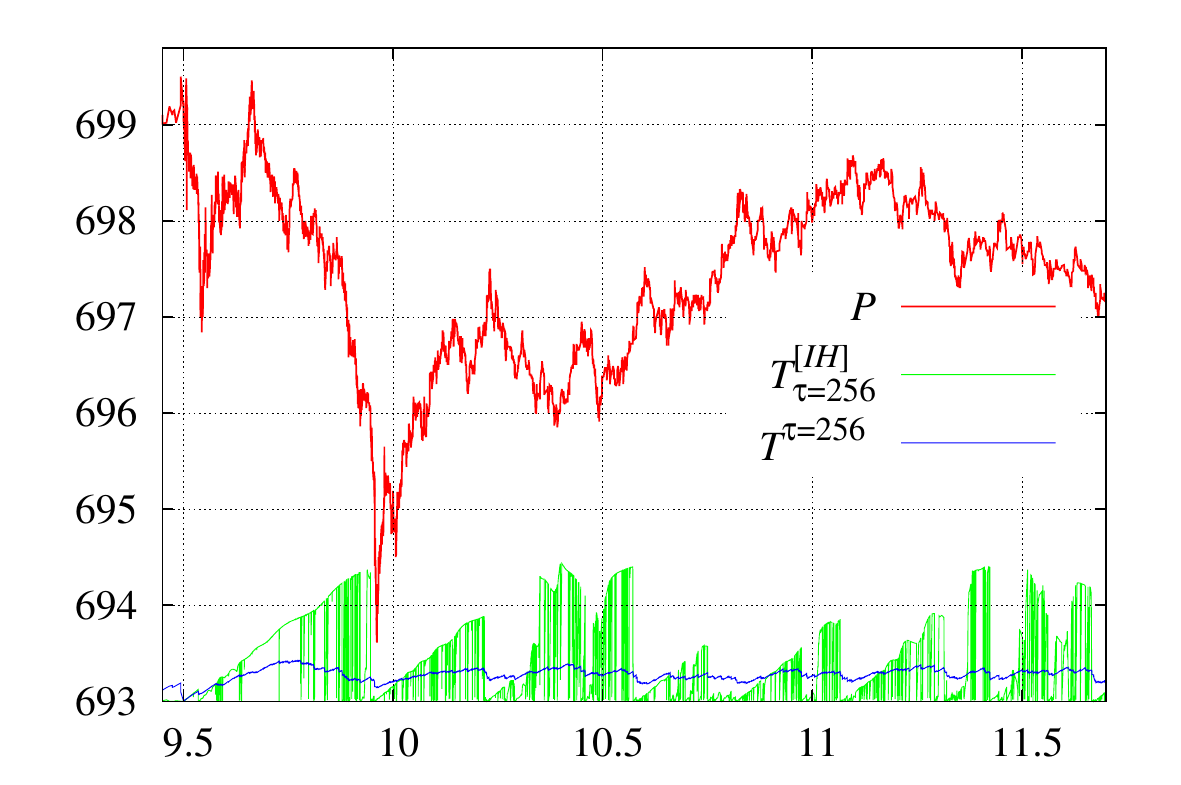}
  \caption{\label{TforImax}
    $T^{[IH]}$ (multiplied by $10^{-3}$ and shifted up to fit the chart)
    and regular moving average $T^{\tau}$
    for $\tau=256$s.
 The AAPL stock on September, 20, 2012.
 The calculations in shifted Legendre basis with $n=12$
    for operator $I=dV/dt$.
  }
\end{figure}
In Fig. \ref{TforImax} the value of $T^{[IH]}$ (scaled by the factor $10^{-3}$
and shifted up to fit the chart) is presented
along with regular moving average $T^{\tau}$
for $\tau=256$s.
One can clearly see that there is no smooth transition
between the states, the ``switch'' happens instantly,
there is no $\tau$-proportional time delay,
what is typical for regular moving averages $T^{\tau}$
and the one in Fig. \ref{MovingAveragePlot}.
A linear dependence of $T^{[IH]}$ on time is also observed, this is an indication
of stability of $\Ket{\psi^{[IH]}}$ state identification.
The value of $T^{[IH]}$ is the time scale; typically it is
easier to work with
the density matrix $\rho_{J(\psi^2)}$
obtained from
$\psi(x)=\psi^{[IH]}(x)$
rather than with the time scale itself.

\section{\label{Conclusion}Conclusion}
A moving average with the weight
$d\mu=dV\,\Phi(x(t))\omega(t)$ is considered.
\begin{align}
  P^{aver}&=\frac{\int_{-\infty}^{t_{now}} dV\,\Phi(x(t))\omega(t)p(t)}{\int_{-\infty}^{t_{now}} dV\,\Phi(x(t))\omega(t)}
  \label{pmovingaverMovingAverage}
\end{align}
The $\omega(t)$ is decaying exponent (\ref{Wbasis});
the polynomial $\Phi(x)$ is obtained solely
from observed execution flow $I=dV/dt$
and has the form $\Phi(x)={\psi^{[IH]}}^2(x)$, it corresponds to
the maximal eigenvalue of (\ref{GEVdef}) eigenproblem.
Contrary to regular moving average (\ref{pmovingaver})
the developed approach has $n-1$ internal degrees of freedom
what adjusts averaging weight according to spikes in
other observable (such as execution flow $I=dV/dt$).
These internal degrees of freedom allow
to obtain
an immediate ``switch'', what is not possible in regular moving average
that always has $\tau$-proportional time delay, lagging indicator.
The comparison with regular moving average
is most clear in Fig. \ref{TforImax}: smooth regular moving average (blue line)
vs. ``switching'' moving average with internal degrees of freedom (green line).

\balance
\bibliographystyle{IEEEtranDOI}
\bibliography{mla,LD,davydov}
\end{document}